\documentclass[conference]{IEEEtran}
\IEEEoverridecommandlockouts
\usepackage{cite}
\usepackage{amsmath}
\usepackage{amssymb}
\usepackage{nicefrac}

\usepackage{tikz}
\usetikzlibrary{arrows}    % siehe arrow tips library
\usepackage{circuitikz}
\usetikzlibrary{circuits.ee.IEC}
\usetikzlibrary{dsp,chains}
\usetikzlibrary{shapes.geometric,calc}

\usepackage{bm}

\usepackage{comment}
\usepackage{siunitx}
\usepackage{epstopdf}
\ifCLASSOPTIONcompsoc
\usepackage[caption=false,font=normalsize,labelfont=sf,textfont=sf]{subfig}
\else
\usepackage[caption=false,font=footnotesize]{subfig}
\fi

\usepackage{changebar}\nochangebars

\usepackage{microtype}

\newcommand{\Kprim}{\bm{K}}
\newcommand{\Kadj}{\tilde{\bm{K}}}
\newcommand{\As}{\bm{\mathcal{A}}}
\newcommand{\Bs}{\bm{\mathcal{B}}}
\newcommand{\Cs}{\bm{\mathcal{C}}}
\newcommand{\Ks}{\bm{\mathcal{K}}}
\newcommand{\Ntx}{N_\textsc{tx}}

\newcommand{\her}{{\text{\tiny$\mathrm{H}$}}}
\newcommand{\tr}{{\text{\tiny$\mathrm{T}$}}}

\usepackage{algorithmic}
\usepackage{graphicx}
\usepackage{textcomp}
\def\BibTeX{{\rm B\kern-.05em{\sc i\kern-.025em b}\kern-.08em
    T\kern-.1667em\lower.7ex\hbox{E}\kern-.125emX}}
\begin{document}

\title{Analytical Models for Particle Diffusion and Flow in a Horizontal Cylinder with a Vertical Force\\
\thanks{\footnotesize This work has been supported by the German Research Foundation (Deutsche Forschungsgemeinschaft, DFG),  grant number RA 801/6-1, SCHO 831/6-1.}
}

\author{\IEEEauthorblockN{Maximilian Sch\"{a}fer, Wayan Wicke, Rudolf Rabenstein, Robert Schober}
\IEEEauthorblockA{\textit{Telecommunications Laboratory (LNT)} \\
\textit{Friedrich-Alexander-Universit\"at Erlangen-N\"urnberg (FAU)}\\
Erlangen, Germany\\
max.schaefer@fau.de, wayan.wicke@fau.de, rudolf.rabenstein@fau.de, robert.schober@fau.de}
}
\IEEEaftertitletext{\vspace{-1\baselineskip}}
\maketitle

\begin{abstract}
This paper considers particle propagation in a cylindrical molecular communication channel, e.g. a simplified model of a blood vessel. Emitted particles are influenced by diffusion, flow, and a vertical force induced e.g.  by gravity or magnetism. The dynamics of the diffusion process are modeled by multi-dimensional transfer functions in a spatio-temporal frequency domain. Realistic boundary conditions are incorporated by the design of a feedback loop. The result is a discrete-time semi-analytical model for the particle concentration in the channel.   
The model is validated by comparison to particle-based simulations. 
These numerical experiments reveal that the particle concentration of the proposed semi-analytical model and the particle-based model are in excellent agreement.
The analytical form of the proposed solution provides several benefits over purely numerical models, e.g. high flexibility, existence of low run-time algorithms, extendability to several kinds of boundary conditions, and analytical connection to parameters from communication theory.
\end{abstract}

%\begin{IEEEkeywords}
%Molecular Communications, Functional Transformations, Diffusion-Advection Problem, Magnetic nanoparticles, Sturm-Liouville Problems
%\end{IEEEkeywords}

\section{Introduction}
The application of communication engineering principles to biotechnological problems has  prompted a large interest and has motivated the investigation of the transport of molecules as carriers of information, see \cite{Farsad_comprehensive_2016} for a review of recent literature.

%Active transport
The transport of molecules, or more generally particles, differs fundamentally from the propagation of electromagnetic waves and therefore requires new theoretical models for system design and analysis \cite{Farsad_comprehensive_2016}.
Though negligible at macroscale, the transport by diffusion becomes relevant at nanoscale.
However, diffusion can be slow and thus relying only on diffusion might be limited to small distances such as in intracellular communication.
In fact, in biotechnology, active transport mechanisms induced by an external force, fluid flow, or a combination thereof are well established \cite{Fournier_Basic_2017}.
%Magnetic nanoparticles
In particular, magnetic forces are attractive because the magnetization of biological particles is often negligible and special purpose magnetic nanoparticles can be engineered and tailored to an application by adapting their size and composition \cite{Pankhurst_Applications_2003}.
In this way, magnetic nanoparticles can be selectively detected and externally steered in a preferred direction by a magnetic field. % \cite{Pankhurst_Applications_2003}.
Furthermore, magnetic nanoparticles can be made biocompatible by suitable coating to avoid reactions with undesired reactants and allow for specific binding to others \cite{Pankhurst_Applications_2003}.

%Magnetic nanoparticles in molecular communication
%Inspired by their versatility in biotechnology, the use of magnetic nanoparticles for molecular communication was studied in \cite{Kisseleff_Magnetic_2017} and \cite{Wicke_Molecular_2017}.
%In particular, \cite{Kisseleff_Magnetic_2017} proposed a wearable device for detecting particles moving through a coil and \cite{Wicke_Molecular_2017} investigated the benefits of using magnetic nanoparticles as attractable information carriers in a two-dimensional flow system.
%Simplified models
%While the focus of \cite{Wicke_Molecular_2017} was in introducing the particles as information carriers, the considered system model was simplistic in considering a two-dimensional environment.
%
%However, investigating the particle transport in a blood vessel is of high interest for molecular communication due to its practicality \cite{Felicetti_Modeling_2014,Chahibi_molecular_2013,He_Channel_2016}.
Motivated by the concept of using magnetic nanoparticles for molecular communication proposed in \cite{Wicke_Molecular_2017, Unterweger_Experimental_2018} and the practicality of studying blood vessels \cite{Felicetti_Modeling_2014,Chahibi_molecular_2013,He_Channel_2016}, here we consider particle transport subject to horizontal fluid flow, diffusion, and a vertical magnetic force in a cylinder model of a blood vessel.
%Thereby, f in axial direction
For simplicity, we assume a uniform axial flow, which is a typical model known as plug flow \cite{Levenspiel_Chemical_1999}. Although blood vessels may have a complicated shape, modeling them as cylinders is well-accepted \cite{Fournier_Basic_2017}. To study the distribution of particles along a segment of the vessel wall, a line receiver model is considered.

When neglecting diffusion, the particle trajectory in a cylinder due to fluid flow and magnetic force can be accurately modeled even for more complicated flow patterns \cite{Furlani_analytical_2006}.
However, modeling diffusion, flow, and particle drift by magnetic force in a cylinder is analytically demanding due to the non-symmetry of the problem.
%Numerical simulation
Therefore, usually finite element numerical simulation is employed to study this type of system, e.g., in drug-targeting \cite{Nacev_behaviors_2011,Afkhami_Ferrofluids_2017}.
%This type of model is also useful for experiments \cite{Janikowska_novel_2017}.
%However, a numerical solution does not give insight into the transport process.
However, numerical solutions lead to algorithms with typically long run-times and low flexibility due to changing channel parameters.

%Our approach: semi-analytic model
%The semi-analytic form is much better than a purely numerical scheme...
A different modeling technique is the application of functional transformations, which are based on the modal expansion of an initial boundary value problem \cite{Curtain:1995:IIL:207416,churchill:1972}. This technique has been already applied successfully, e.g. in the field of sound synthesis and circuit modeling, see \cite{lnt2017-47,antonini2012} for further references. 
This kind of modeling technique offers several benefits: It leads to semi-analytical models in terms of a state-space description which provide insights into the diffusion process.
%from an engineering point of view
Furthermore, these models allow the adaptation of their boundary behavior using feedback techniques \cite{lnt2017-47,Deutscher:IJC:2009}.
%Applicability

%Our contribution
%The proposed framework is extensible to different boundary conditions and ...
%The used algorithm is fast ...
%In this paper, a semi-analytical model for the diffusion process in a cylinder with a vertical magnetic force is derived in terms of a state space description.   

This paper proposes a semi-analytical model for a diffusive cylindrical molecular communication channel in the presence of a magnetic force and flow. The model leads to a low run-time simulation algorithm, which provides several benefits compared to purely numerical models. 
%\cbstart
The derivation of the model follows the mathematically more detailed description in \cite{schaefer:dsp:2018}, where particle diffusion under the influence of a magnetic force in a circular disk is simulated.  
%\cbend
%
%
%This leads to a discrete-time simulation algorithm for the spatial concentration of particles in a receiver.
%The derived model can also serve as a blueprint for scenarios with different boundary conditions. 
Our results might be useful for studying molecular communication in, e.g. blood vessels, flow reactors or microfluidic channels.  It is also applicable to the study of diffusion with horizontal slurry flow in tubes where particles sediment due to gravity \cite[Chapter~13]{brennen2005fundamentals}.
%The validity of the model is shown by simulation and comparison with particle-based simulation results. 

%Paper structure
The paper is structured as follows: In Section~\ref{sec:prob}, a physical description of the scenario under study is presented.
%in terms of an initial-boundary-value problem
%, which is then formulated in a unifying vector form
In Section~\ref{sec:model}, a semi-analytical model for the diffusive cylindrical channel is derived and a simulation algorithm is proposed. The validity of the model is shown in Section~\ref{sec:experiments} by simulation, and the algorithm is analyzed and its benefits are discussed. Finally, Section~\ref{sec:conclusion} concludes the paper and proposes topics for future work.  

%%%% Problem Description %%%% 
\section{Problem Description}
\label{sec:prob}

\begin{figure}
	\begin{tikzpicture}
%	[auto, scale = 0.9, every node/.style={scale=0.7}, font=\LARGE, node distance=2.5cm,>=latex', 
%	rounded corners=2pt]
%			\draw[help lines] (6,6) grid (-3,-3);
	\node[cylinder,draw=black,thick,aspect=3.5,minimum height=5cm,minimum width=1.5cm,shape border rotate=0] (A) {};
	
	\draw[dashed]
	let \p1 = ($ (A.after bottom) - (A.before bottom) $),
	\n1 = {0.5*veclen(\x1,\y1)-\pgflinewidth},
	\p2 = ($ (A.bottom) - (A.after bottom)!.5!(A.before bottom) $),
	\n2 = {veclen(\x2,\y2)-\pgflinewidth}
	in
	([xshift=-\pgflinewidth] A.before bottom) arc [thick, start angle=270, delta angle=180,
	x radius=\n2, y radius=\n1];
	
	\node[circle,inner sep=0pt, minimum size=0.2cm, draw,fill=red!30](part) at (-1,0.2){};
	\draw[dspconn] (part) to (-1,-0.6);
	\draw[dspconn] (part) to (0.5,0.2);
	
	\node at (-0.25,0.4) {
		\parbox{1cm}{
			\centering $v_z'$
	}};
	\node at (-0.75,-0.35) {
		\parbox{1cm}{
			\centering $u'$
	}};
	
	\draw[dspline,thick] (-2,-0.8) to (-2,-1.1);
	\draw[dspconn,thick] (-2,-0.95) to (2.7,-0.95);
	\node at (-2,-1.2) {
		\parbox{1cm}{
			\centering $z' = 0$
	}};
	\node at (2.7,-1.2) {
		\parbox{1.5cm}{
			\centering $z' \to \infty$
	}};
	
	%Coordinate System
	\draw[thick] (4.5,0) circle (0.75cm);
	\draw[dspconn,thick] (3.55,0) -- (5.8,0);
	\draw[dspconn,thick] (4.5,-0.95) -- (4.5, 1.2);
	
	%Emitter points
	\draw[fill=blue!30] (4.125,0) circle (0.1cm);
	\draw[fill=blue!30] (-2,-0.2) circle (0.08cm);
	
	% Axis Labels 
		\node at (5.8,-0.2) {
			\parbox{1cm}{
				\centering $x'$
		}};
		\node at (4.7,1.2) {
			\parbox{1cm}{
				\centering $y'$
		}};
		\node at (5.38,-0.15) {
			\parbox{1cm}{
				\centering $a'$
		}};

	% Emitter Label
	\node at (2.5,-1.55) {
		\parbox{1cm}{
			\centering \small\mbox{Emitter at $x_0' = -0.5a'$, $y_0' = 0$}
	}};
	\draw[dspconn] (3.4,-1.33) -- (4,-0.1);
	
	%u in right ploot
	\draw[dspconn] (4.125, -0.1) to (4.125, -0.6);

	\node at (4.3,-0.3) {
		\parbox{1cm}{
			\centering \small\mbox{$u'$}
	}};

	% Line Receivers
	\draw[dspline,thick] (0.5,-0.3) -- (1.4,-0.3);
	\draw[dspline,thick] (0.8,-0.6) -- (1.7,-0.6);
	
	\node at (0.9,-0.13) {
		\parbox{1cm}{
			\centering $d'$
	}};
	
	\node at (-0.5,-1.55) {
		\parbox{1cm}{
			\centering \small\mbox{Line Receivers}
	}};

	\draw[dspconn] (0,-1.33) -- (0.55,-0.38);
	\draw[dspconn] (0,-1.33) -- (0.85,-0.68);
	
	\end{tikzpicture}
%	\vspace*{-3ex}
    \caption{Diffusing particles (red circle) in a cylinder subject to a horizontal flow $v_z'$ and a vertical velocity $u'$. The cylinder is infinitely extended in $z'$-direction. Particles are emitted by a point source (blue circle) and detected by line receivers of length $d'$.}
    \label{fig:1}
%    \vspace*{-3ex}
\end{figure}
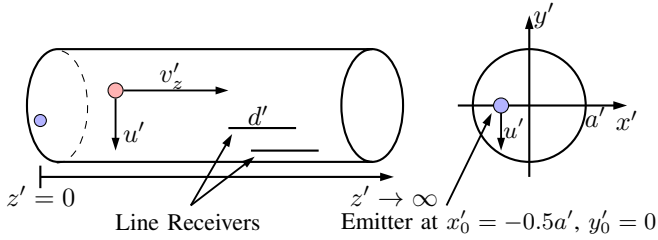

Fig.~\ref{fig:1} shows a simplified cylindrical model of a blood vessel. Information carrying nano particles are emitted into the channel of radius $a'$ by a point source at $z' = 0, x_0', y_0'$. They are diffusing with a constant diffusion coefficient $D'$ and are dragged by a uniform flow in $z'$-direction with velocity $v_z'$. The particle movement is controlled by a vertical magnetic force causing a constant uniform drift velocity $u'$, see \cite{Wicke_Molecular_2017} for details. For modeling the reception, the particle concentration is desired at a point on the vessel wall where a receiver nanomachine is mounted. As an example generalization, for probing the dynamics of the particle concentration over an extended axial distance, a line receiver of length $d'$ is considered. The particle concentration over time in the receiver is denoted by $p_{\mathrm{r},z'}'(x',y',t')$ (the $z'$-index indicates the position in $z'$-direction), the particle flux is denoted by vector $\bm{i}'$.

%In Fig.~\ref{fig:1}, the scenario under study is shown. Particles emitted in the cylinder with radius $a'$ diffusing with a constant diffusion coefficient $D'$ are dragged by a uniform flow in $z'$-direction with the velocity $v_z'$. 
%Additionally a vertical magnetic force affects the particles causing a constant uniform drift velocity $u'$. 
%Particles are emitted by a point source at $r_0', \varphi_0'$ and detected by elongated receivers of length $d'$. 

%\vspace*{-2ex}
\subsection{Normalization}
\label{subsec:norm}
Before the physical description of the scenario in Fig.~\ref{fig:1} in terms of partial differential equations (PDEs) is established, the variables in the system are normalized. Introducing a reference time $\tau = \frac{a'^2}{D'}$ and a reference length $\lambda = a'$ in terms of the radius $a'$ of the cylinder and the diffusion coefficient $D'$ leads to the following normalization 
\begin{align}
& (x,y,z) = \frac{(x',y',z')}{\lambda}, &\bm{i} = \bm{i}'\cdot\lambda^2 \tau, &&p =  p'\cdot\lambda^3, &&& 
a = \frac{a'}{\lambda},
\end{align}
%\vspace*{-3ex}
\begin{align}
&D = D' \cdot \frac{\tau}{\lambda^2}, \,\,\,& v_z = v_z' \cdot\frac{\tau}{\lambda}, && u = u' \cdot\frac{\tau}{\lambda}, &&&t = t' \cdot\frac{1}{\tau}.
\end{align}
In the following, concentrations and fluxes in the cylindrical geometry are formulated in polar coordinates, using the mapping $x=r\cos(\varphi)$ and $y=r\sin(\varphi)$, particularly $y_0 = r_0\sin(\varphi_0)$.
With this normalization and substitution a normalized dimensionless physical model is considered in the following sections. 

\vspace*{2ex}
\subsection{Physical Model}
\label{subsec:physic}
%The complete scenario can be described by a PDE including boundary and initial conditions. 
%
Due to the geometry of the problem, the average concentration in a line receiver, $p_{\mathrm{r},z}$, following a point release, can be expressed as
%\vspace*{-1.1ex}
\begin{align}
p_{\mathrm{r},z}(r,\varphi,t)=p_{xy}(r,\varphi,t)\cdot p_{d,z}(t). \label{eq:p0}
\end{align}
Here, $p_{d,z}$ is the accumulated concentration in the line segment for a one-dimensional drift-diffusion process in $z$-direction. It is obtained by integration \cite[Eq.~(23)]{Farsad_comprehensive_2016}
%\vspace*{-1ex}
\begin{align}
p_{d,z}(t) = \int_{z-\frac{d}{2}}^{z + \frac{d}{2}}  \frac{1}{\sqrt{4\pi t}} \exp\Big(-\frac{(\tilde{z}-v_zt)^2}{4t}\Big)\,\mathrm{d}\tilde{z}. \label{eq:p1a}
\end{align}
Furthermore, $p_{xy}$ is the concentration of a diffusion process in a circular cross-section of a cylinder. Its calculation is the main task in the following sections. We note that a similar separation as in \eqref{eq:p0} also holds for the flux $\bm{i}$.

The particle concentration $p_{xy}$ in the circular disk $\varphi\in[0,2\pi)$, \mbox{$r\in[0,a]$} under the influence of the vertical drift velocity $u$ is given by the following PDE
\begin{IEEEeqnarray}{rCl+l}
    \label{eq:original_problem}
    \frac{\partial}{\partial t} p_{xy}(r,\varphi,t) &=& -\mbox{\textbf{div}}\, \bm{i}_{xy}(r,\varphi,t) & \hspace*{-14ex}r\leq a, t>0 \IEEEyesnumber\IEEEyessubnumber* \label{eq:p3}\\
    \bm{i}_{xy}(r,\varphi,t) &=& -\mbox{\textbf{grad}}\, p_{xy}(r,\varphi,t) - \bm{e}_y u p_{xy}(r,\varphi,t) &  \label{eq:p4}\\
    i_r(r,\varphi,t) &=& 0 & \hspace*{-8ex} r=a\label{eq:p5}\\
    p_{xy}(r,\varphi,t) &=& \frac{1}{r_0}\delta(r-r_0)\delta(\varphi-\varphi_0) & \hspace*{-8ex} t=0, \label{eq:p6}
\end{IEEEeqnarray}
\cbstart
with the unit vector in $y$-direction $\bm{e}_y$, the delta impulse $\delta(\cdot)$ and gradient $\mbox{\textbf{grad}}$ and divergence $\mbox{\textbf{div}}$ in cylindrical coordinates. 
\cbend
The emission of particles is described by the initial condition \eqref{eq:p6} and the zero flux at $r = a$ by the boundary condition \eqref{eq:p5}.  
%The steady state solution of this PDE for $ t \to \infty$ can be calculated as $p(r,\varphi,t\to\infty) = \frac{u}{2\pi I_1(au)}e^{-uy}$, where $I_1$ is the first order modified Bessel function.    

%\vspace*{-2ex}
\subsection{Transformation of Variables}
\label{subsec:tov}
The presence of the drift term in \eqref{eq:p4} highly scales the complexity of a direct solution of the system \eqref{eq:p3}, \eqref{eq:p4}. Therefore, the problem is transformed into an auxiliary problem with the auxiliary particle concentration $q_{xy}$ by application of the substitution $p_{xy}(r,\varphi,t)=q_{xy}(r,\varphi,t)\exp(-\frac{u}{2}(y-y_0)-\frac{u^2}{4}t)$, see also \cite{Perez_transform_2009}
\begin{IEEEeqnarray}{rCl+l}
    \frac{\partial}{\partial t} q_{xy}(r,\varphi,t) &=& -\mbox{\textbf{div}}\, \bm{i}_{xy}(r,\varphi,t) & \hspace*{-6ex} r\leq a, t>0 \IEEEyesnumber\IEEEyessubnumber* \label{eq:p7}\\
    \bm{i}_{xy}(r,\varphi,t) &=& -\mbox{\textbf{grad}}\, q_{xy}(r,\varphi,t) &  \label{eq:p8}\\
    i_r(r,\varphi,t) &=& \frac{u}{2}\sin\varphi \cdot q_{xy}(r,\varphi,t) & r=a\label{eq:p9}\\
    q_{xy}(r,\varphi,t) &=& \frac{1}{r_0}\delta(r-r_0)\delta(\varphi-\varphi_0) & t=0. \label{eq:p10} 
\end{IEEEeqnarray}
The main task in the following sections is to derive a semi-analytical model for the auxiliary concentration $q_{xy}$ with the non-trivial boundary condition \eqref{eq:p9}. In the end, this result is used to obtain a semi-analytical simulation model for the desired concentration \eqref{eq:p0} at the receiver.

%%%% Vector Formulation %%%% 
%\vspace*{-1.8ex}
\subsection{Vector Formulation}
\label{subsec:vector}
Before the transfer function model for the considered diffusion process is constructed in Section~\ref{sec:model}, the auxiliary PDE in \eqref{eq:p7}, \eqref{eq:p8}, the boundary conditions \eqref{eq:p9}, and the initial conditions \eqref{eq:p10} are rearranged into a unified vector form according to \cite{lnt2017-47}. 
%\vspace*{-1ex}
%\subsubsection{Partial Differential Equation}
The reformulation of the PDE in \eqref{eq:p7}, \eqref{eq:p8} in terms of a matrix valued spatial differential operator $\bm{L}$ defined on the circular disk $\bm{x} = \left[r, \varphi \right]$ with $r \in \left[0, a\right]$ and $\varphi \in \left[-\pi, \pi\right)$ leads to 
\begin{align}
&\left[\frac{\partial}{\partial t}\bm{C} - \bm{L} \right]\bm{y}(\bm{x},t) = \bm{0}, &\bm{L} = \bm{A} + \nabla \bm{I},
\label{eq:1}
\end{align}
with a mass matrix $\bm{C}$, a matrix of damping parameters $\bm{A}$, and the identity matrix $\bm{I}$. The involved block matrices and operators are defined as
\begin{align}
&\bm{A} = \begin{bmatrix}
0 & -\bm{I}\\
0 & \bm{0}
\end{bmatrix}, 
&\!\!\!\bm{C} = \begin{bmatrix}
0 & \bm{0} \\
1 & \bm{0}
\end{bmatrix}, 
&& \!\!\!\nabla = 
\begin{bmatrix}
- \mbox{\textbf{grad}} & \bm{0}\\
0 & -\mbox{\textbf{div}}
\end{bmatrix}.
\label{eq:2}
\end{align}
Vector $\bm{y}$ contains the independent variables of the system: the auxiliary particle concentration and the flux in $r$- and $\varphi$-direction 
\begin{align}
\begin{split}
\bm{y}(\bm{x},t) &= \begin{bmatrix}
q_{xy}(\bm{x},t) & \!\!\!\!\bm{i}_{xy}^\tr(\bm{x},t)
\end{bmatrix}^\tr, \\
\bm{i}_{xy}(\bm{x},t) &= \begin{bmatrix}
i_r(\bm{x},t) & \!\!\!\!i_\varphi(\bm{x},t) 
\end{bmatrix}^\tr\!\!,
\end{split}
\label{eq:3}
\end{align}
where $()^\tr$ denotes the transposed.
According to the dimensions of $\bm{i}_{xy}$ and $q_{xy}$ in \eqref{eq:3}, the matrices and vectors in \eqref{eq:2} have sizes $3\times3$ and $3\times1$, respectively.
%
%\subsubsection{Initial and Boundary Conditions}
The initial conditions for the auxiliary particle concentration $q_{xy}$ in \eqref{eq:p10} are rearranged similar to the vector of variables in \eqref{eq:3}
%\vspace{-1ex}
\begin{align}
\bm{y}_\mathrm{i}(\bm{x}) = \begin{bmatrix}
q_{xy}(\bm{x}, 0) & 0 & 0 
\end{bmatrix}^\mathrm{T}. 
\label{eq:4}
\end{align}
In addition to the desired boundary behavior \eqref{eq:p9}, a second boundary condition is defined 
%\vspace{-1ex}
\begin{align}
i_r(a,\varphi,t) = \phi(\varphi,t),
\label{eq:6}
\end{align}
with a general boundary excitation $\phi$. 
The second set of conditions \eqref{eq:6} is used to design a general transfer function model for the diffusion process in Section~\ref{subsec:ftm}. The derived model serves as a blueprint for a diffusion process, where different kinds of boundary conditions can be incorporated. In Section~\ref{subsec:fcl}, the desired boundary behavior \eqref{eq:p9} is realized by a modification of the blueprint model.

\subsection{Laplace Transformation}
Applying the Laplace transform to the PDE in \eqref{eq:1} and \eqref{eq:4} leads to 
%\vspace*{-1ex}
\begin{align}
\left[s\bm{C} - \bm{L}\right]\bm{Y}(\bm{x},s) = \bm{C}\bm{y}_\mathrm{i}(\bm{x}),
\label{eq:7}
\end{align}
where $\bm{Y}$ is the frequency domain equivalent of the vector of variables $\bm{y}$ in \eqref{eq:3}. The complex frequency is denoted by $s$. The frequency domain representation of the PDE in \eqref{eq:7} and the boundary conditions \eqref{eq:p9}, \eqref{eq:6} are the basis for the subsequent transformations. 

%This boundary condition can be formulated to fit into the vector form with the vector of variables $\bm{y}$ with a matrix $\bm{F}_\mathrm{c}$ transforming the vector of variables  
%\begin{align}
%&\bm{F}_\mathrm{c}^\mathrm{H}\bm{y}(\bm{x},t) = \begin{bmatrix}
%0 & \bm{0}^\mathrm{T} \\
%\bm{f}_{\mathrm{c}1}^\mathrm{H} & \bm{F}_{\mathrm{c}2}^\mathrm{H}
%\end{bmatrix} \cdot 
%\begin{bmatrix}
%q(\bm{x},t) \\
%\bm{i}(\bm{x},t)
%\end{bmatrix} = 
%\bm{0}, &r = a, 
%\label{eq:6}
%\end{align}
%with the matrices 
%\begin{align}
%&\bm{F}_{\mathrm{c}2}^\mathrm{H} = \begin{bmatrix}
%1 & 0 \\
%0 & 0 
%\end{bmatrix}, 
%& \bm{f}_{\mathrm{c}1}^\mathrm{H} = \begin{bmatrix}
%-\frac{u}{2}\sin \varphi \\
%0
%\end{bmatrix}. 
%\end{align}

%%%% Model %%%% 
\section{Semi-Analytical Model}
\label{sec:model}

In this section, a transfer function model for the circular diffusion process is derived. First, a model for the process described by the PDE in \eqref{eq:7} with the second set of boundary conditions in \eqref{eq:6} is established using a modal expansion of the spatial differential operator $\bm{L}$ into eigenfunctions \cite{Curtain:1995:IIL:207416,churchill:1972}. Then, this model is formulated as a state-space description and is forced to fulfill the boundary conditions in \eqref{eq:p9} by the design of a feedback loop \cite{Deutscher:IJC:2009}.

\subsection{Transfer Function Model}
\label{subsec:ftm}
To derive a transfer function for the given 2-D diffusion process \eqref{eq:7}, the spatial differential operator $\bm{L}$ is expanded into eigenfunctions $\Kadj$ (see Section \ref{subsubsec:eig}) based on the Sturm-Liouville theory \cite{churchill:1972}.  

\subsubsection{Sturm-Liouville Transformation}
For the transformation of space variables, a Sturm-Liouville transformation is defined in terms of the integral transform \cite{churchill:1972}
%\vspace*{-1ex}
\begin{align}
\mathcal{T}\left\lbrace \bm{Y}(\bm{x},s) \right\rbrace \!= \!\bar{Y}_n(\mu,s) \!=\!\! \int_{0}^{2\pi} \!\! \int_{0}^{a} \!\!\!\tilde{\bm{K}}_n^\her(\bm{x},\mu) \bm{C}\bm{Y}(\bm{x},s) \,r\,\mathrm{d}r \,\mathrm{d}\varphi,
\label{eq:8}
\end{align}
with the eigenfunctions $\Kadj_n$. The inverse transformation is a generalized Fourier series with the eigenfunctions $\Kprim_n$ and the scaling factor $N_{\mu,n}$
%\vspace*{-2ex}
\begin{align}
\mathcal{T}^{-1}\!\!\left\lbrace \bar{Y}_n(\mu, s) \right\rbrace \!=\! \bm{Y}(\bm{x},s) \!= \!\!\!\!\!\sum_{n=-\infty}^\infty \sum_{\mu = 0}^\infty \frac{1}{N_{\mu,n}} \bar{Y}_n(\mu,s) \bm{K}_n(\bm{x},\mu),
\label{eq:9}	
\end{align} 
where $\bar{Y}_n(\mu,s)$ is a scalar representation of the vector of variables in \eqref{eq:3} in the temporal and spatial transform domain. Index $\mu \in \mathbb{Z}$ is the index of a spatial frequency variable $s_{\mu,n}$ for which \eqref{eq:1} has nontrivial solutions \cite{churchill:1972,lnt2017-47}. 

Both sets of eigenfunctions $\Kprim_n$ and $\Kadj_n$ are determined from problem-specific eigenvalue problems \cite{lnt2017-47}. The index $n \in \mathbb{Z}$ is here already introduced because of the Bessel-nature of the eigenfunctions, cf. Section~\ref{subsubsec:eig}.

\subsubsection{Transform Domain Representation}
Applying the forward transform \eqref{eq:8} to the PDE \eqref{eq:7} and exploiting the properties of the eigenfunctions (see \cite{lnt2017-47,churchill:1972}) leads to 
\begin{align}
s\bar{Y}_n(\mu,s) - s_{\mu,n}\bar{Y}_n(\mu,s) = \bar{y}_{\mathrm{i},n}(\mu) + \bar{\Phi}_{n}(\mu,s), 
\label{eq:10} 
\end{align}
with the transform domain representation of the initial conditions $\bar{y}_{\mathrm{i},n}(\mu)$ from \eqref{eq:4} and the transformed boundary term $\bar{\Phi}_{n}(\mu,s)$ given by
%\vspace*{-1ex}
\begin{align}
\bar{y}_{\mathrm{i},n}(\mu) &= \int_{0}^{2\pi} \!\!\!\int_{0}^{a} J_n(k_{\mu,n}r)\,e^{-jn\varphi} q_{xy}(\bm{x},0) \,r\,\mathrm{d}r \,\mathrm{d}\varphi,\label{eq:i1}\\
\bar{\Phi}_{n}(\mu,s) &= -\int_{0}^{2\pi} a J_n(k_{\mu,n}a)\,e^{-jn\varphi}\, \Phi(\varphi,s)\, \mathrm{d}\varphi, 
\label{eq:i2}
\end{align}
with the $n$-th order Bessel functions of the first kind $J_n$ and $\Phi$ is the frequency domain equivalent of the boundary excitation $\phi$ in \eqref{eq:6}. The values $s_{\mu,n} = -k^2_{\mu,n}$ represent the spatial eigenfrequencies of the diffusion process. Solving \eqref{eq:10} for the transformed vector of variables leads to a representation in terms of a multidimensional transfer function
%\vspace*{-2.5ex} 
\begin{align}
&\bar{Y}_n(\mu,s) = \frac{1}{s - s_{\mu,n}} \left( \bar{y}_{\mathrm{i},n}(\mu) + \bar{\Phi}_{n}(\mu,s) \right), &\Re\{s_{\mu,n}\} < 0.
\label{eq:11}
\end{align}
%\vspace*{-2.5ex}
\subsubsection{Eigenfunctions and Eigenfrequencies}
\label{subsubsec:eig}
The eigenfunctions for the forward and inverse transformation are derived by solving a dedicated eigenvalue problem \cite{churchill:1972}. Its derivation is skipped here for brevity but it is described for similar problems in \cite{lnt2017-27,lnt2017-47}. The eigenfunctions in \eqref{eq:8}, \eqref{eq:9} are obtained as
\begin{align}
\begin{split}
\Kprim_n(\bm{x},\mu) &= \begin{bmatrix}
J_n(k_{\mu,n} r)\\
-k_{\mu,n}J'_n(k_{\mu,n}r)\\
-\frac{1}{r}(jn)J_n(k_{\mu,n}r)
\end{bmatrix} e^{jn\varphi},\\
\Kadj_n(\bm{x},\mu) &= \begin{bmatrix}
k_{\mu,n}J'_n(k_{\mu,n}r)\\
\frac{1}{r}(jn)J_n(k_{\mu,n}r)\\
J_n(k_{\mu,n} r)
\end{bmatrix} e^{jn\varphi},
\end{split}
\label{eq:12}
\end{align}
where $J'(x) = \nicefrac{\partial}{\partial x}J(x)$. The eigenfrequencies $s_\mu = -k^2_{\mu,n}$ of the diffusion process are calculated by the evaluation of boundary conditions similar to \eqref{eq:6}, i.e., from the real-valued zeros of $J'_n(k_{\mu,n}a) = 0$.

Due to the bi-orthogonality of the eigenfunctions, the scaling factor $N_{\mu,n}$ is calculated by evaluation of the  integral \cite{churchill:1972}
%\vspace*{-1ex}
\begin{align}
N_{\mu,n} &= \int_{0}^{2\pi} \!\!\!\int_{0}^{a}\Kadj_n^\her(\bm{x},\mu) \bm{C} \Kprim_n(\bm{x},\mu) \,r\,\mathrm{d}r \,\mathrm{d}\varphi.
\label{eq:14}
\end{align}

\subsubsection{State-Space Description}
\label{subsubsec:ssd}
Grouping all elements of the matrices and vectors in \eqref{eq:10} in the range of $n, \mu$ into vectors gives a state equation in the continuous frequency domain
\begin{align}
s\bar{\bm{Y}}(s) = \As \bar{\bm{Y}}(s) + \bar{\bm{y}}_\mathrm{i} + \bar{\bm{\Phi}}(s), 
\label{eq:15}
\end{align}
with the matrices and vectors 
\begin{align}
&\bar{\bm{Y}}(s) = \begin{bmatrix}
\dots, \bar{Y}_n(\mu,s), \dots 
\end{bmatrix}^\tr, 
%&\bar{\bm{Y}}(s) = \begin{bmatrix}
%\bar{Y}_{-\infty}(0,s), \dots, \bar{Y}_{\infty}(\infty,s),  
%\end{bmatrix}^\tr,
&&\bar{\bm{y}}_\mathrm{i}= \begin{bmatrix}
\dots, \bar{y}_{\mathrm{i},n}(\mu), \dots
\end{bmatrix}^\tr, \label{eq:17}\\%\label{eq:16}
&\bar{\bm{\Phi}}(s) = \begin{bmatrix}
\dots, \bar{\Phi}_{n}(\mu,s), \dots 
\end{bmatrix}^\tr, 
&&\As = \mathrm{diag}\left(\dots, s_{\mu,n}, \dots\right). \label{eq:19} %\label{eq:18}
\end{align}
The diagonal matrix $\As$ contains the eigenvalues $s_{\mu,n}$ of the system on its main diagonal.
The reformulation of the inverse transformation in \eqref{eq:9} leads to an output equation for the state-space description,
%\vspace*{-2ex}
\begin{align}
&\bm{Y}(\bm{x},s) = \Cs(\bm{x}) \bar{\bm{Y}}(s),
\label{eq:20}
\end{align}
where 
%\vspace{-2ex}
\begin{align}
\Cs(\bm{x}) = 
\begin{bmatrix}
\bm{c}_q^\tr(\bm{x})\\
\bm{c}_r^\tr(\bm{x})\\
\bm{c}_\varphi^\tr(\bm{x})
\end{bmatrix}
= \begin{bmatrix}
\dots,\, \frac{1}{N_{\mu,n}}
\begin{bmatrix}
K_{1,n}(\bm{x},\mu)\\
K_{2,n}(\bm{x},\mu)\\
K_{3,n}(\bm{x},\mu)
\end{bmatrix}
, \,\dots 
\end{bmatrix}.
\label{eq:21}
\end{align}
Here, $K_{m,n}$ is the $m$-th element of eigenfunction $\bm{K}_n$ in \eqref{eq:12}. 
The complete state-space description based on state equation \eqref{eq:15} and output equation \eqref{eq:20} is shown in Fig.~\ref{fig:3} (switch open). The matrices and vectors in \eqref{eq:15}, \eqref{eq:20} are theoretically of infinite size (see range of $n,\mu$ in \eqref{eq:9}), and therefore they have to be interpreted in an operator sense. For numerical evaluation, the sums in \eqref{eq:9} have to be truncated so that the matrices become finite and computable. 	  

This state-space model was obtained for the boundary conditions in \eqref{eq:6} and includes the yet undetermined general boundary term $\bar{\bm{\Phi}}(s)$. Setting this term to zero corresponds to the scenario of particle diffusion in a cylinder with zero flux at the boundary \cite[p. 378]{carslaw:1946}. In the next section, the boundary terms are used to realize the desired boundary behavior \eqref{eq:p9}. 

%\vspace*{-1.5ex}
\subsection{Feedback Loop}
\label{subsec:fcl}
In this section, a feedback loop is designed to adjust the boundary behavior of the state-space model from Section~\ref{subsubsec:ssd}, so that it fulfills the boundary conditions in \eqref{eq:p9} \cite{Deutscher:IJC:2009,Willems:ICS:2007:4384643,lnt2017-47}. 
%\vspace*{-0.5ex}
\subsubsection{Feedback Matrix}
The state-space model in \eqref{eq:15}, \eqref{eq:20} is designed with the general boundary term $\bar{\Phi}_n(\mu,s)$. Now, the excitations $\phi$ are replaced by the desired ones \eqref{eq:p9}. The goal is an expression for $\bar{\Phi}_n(\mu,s)$ in terms of the state vector $\bar{\bm{Y}}(s)$ of the state-space description.
A feedback matrix for the system is constructed according to the concepts described in \cite{lnt2017-47}. The starting point is the general transformed boundary term \eqref{eq:i2}.
%\begin{align}
%\bar{\Phi}_n(\mu,s) = - \int_{0}^{2\pi}a K^*_{3,n}(a,\varphi,\mu) \,\Phi_{\mathrm{s}}(\varphi,s) \mathrm{d}\varphi.
%\label{eq:22} 
%\end{align}
Inserting the desired boundary conditions \eqref{eq:p9} into the general term $\phi$ in \eqref{eq:6} (in the frequency domain) leads to
\begin{align}
I_r(a,\varphi,s) = \Phi(\varphi,s) = \frac{u}{2}\sin\varphi\cdot Y_1(a,\varphi,s).
\label{eq:23}
\end{align}
This representation is inserted into the transformed boundary term \eqref{eq:i2}. Exploiting the output equation \eqref{eq:20} and \eqref{eq:23} leads to 
\begin{align}
\bar{\Phi}_n(\mu,s) &= - a J_n(k_{\mu,n}a) \int_{0}^{2\pi} \!\!\!e^{-jn\varphi} \frac{u}{2}\sin\varphi \,\,\bm{c}_q^\tr(a,\varphi) \bar{\bm{Y}}(s) \mathrm{d}\varphi.
\label{eq:24}
\end{align}
The integration over $\varphi$ and considering \eqref{eq:21}, \eqref{eq:12}, and \eqref{eq:9} leads to  
\begin{align}
\bar{\Phi}_n(\mu,s) \!=\! - \hat{b}_n(\mu) \frac{u}{2} \sum_{m=-\infty}^\infty \sum_{\nu = 0}^\infty \frac{J_m(k_{\nu,m}a) f(m,n)}{N_{\nu,m}}  \bar{Y}_m(\nu,s),\label{eq:25}
\end{align}
with the coefficients $\hat{b}_n(\mu) = a J_n(k_{\mu,n}a)$ and a function $f$ defined in terms of Kronecker delta functions and the imaginary unit $j$
\begin{align}
f(m,n) = \frac{\pi}{j}\left(\delta_{m,(n-1)} - \delta_{m,(n+1)}\right). 
\label{eq:26}
\end{align}
Sorting the sums in \eqref{eq:25} into vectors leads to an expression for the transformed boundary term realizing the realistic boundary conditions in terms of the state vector
\begin{align}
\bar{\bm{\Phi}}(s) = - \Bs \Ks \bar{\bm{Y}}(s), 
\label{eq:27}
\end{align}
with the feedback matrix $\Bs \Ks$ 
\begin{align}
&\Bs \Ks = \begin{bmatrix}
\vdots \\
\hat{\bm{b}}_n \hat{\bm{k}}_n^\tr\\
\vdots
\end{bmatrix},
&\hat{\bm{b}}_n = \begin{bmatrix}
\vdots\\
\hat{b}_n(\mu) \\
\vdots
\end{bmatrix},
&&\hat{\bm{k}}_n^\tr &= \begin{bmatrix}
\dots, \hat{\bm{k}}_{m,n}^\tr,\dots
\end{bmatrix},
\label{eq:28} 
\end{align}
%\vspace*{-1.3ex}
\begin{align}
\hat{\bm{k}}_{m,n}^\tr &= \begin{bmatrix}
\dots, \frac{1}{N_{\nu,m}}J_m(k_{\nu,m} a) \frac{u}{2}\delta(m,n),\dots
\end{bmatrix}.	 \label{eq:29}
\end{align}
The system shown in Fig.~\ref{fig:3} with open switch realizes the diffusion process \eqref{eq:p7}, \eqref{eq:p8} with boundary conditions \eqref{eq:6}. By incorporation of \eqref{eq:27} and closing the switch, the system fulfills the desired boundary conditions \eqref{eq:p9}.

\subsubsection{Modified State-Space Description}
\label{subsubsec:mssd}

\begin{figure}[t]
	\centering
	\begin{tikzpicture}[auto, scale = 0.8, every node/.style={scale=0.8}, font=\large, node distance=2.5cm,>=latex', 
	rounded corners=2pt]
	%%%%%%%%%%%%%%%%%%%%%%%%%%%%%%%%%%%%%%%%
%	\draw[help lines] (0,0) grid (15,5);
	\node[draw, dspsquare, minimum width=1cm, minimum height=1cm](s)at (8, 4) {
		\parbox{1cm}{
			\centering \LARGE\mbox{$\nicefrac{1}{s}$}
	}};
	\node[draw, dspsquare, minimum width=1cm, minimum height=1cm](a)at (8, 2.5) {
		\parbox{1cm}{
			\centering $\bm{\mathcal{A}}$
	}};
	\node[draw,dspadder](add) at (6,4){};
	\node[draw,fill](p) at (10,4){};
	\node[draw, dspsquare, minimum width=1cm, minimum height=1cm](c)at (12, 4) {
		\parbox{1cm}{
			\centering $\bm{\mathcal{C}}(\bm{x})$
	}};
	\draw[dspflow,double] (4,4) node[dspnodeopen]{$\bar{\bm{\Phi}}(s)$} to (add);
	\draw[dspline,double] (add) to (s) to (p);
	\draw[dspline,double] (p) to (10,2.5) to (a);
	\draw[dspflow,double] (a) to (6,2.5); 
	\draw[dspline,double] (6,2.5) to (add);
	\draw[dspline,double] (p) to (c);
	\draw[dspline,double] (c) to (14,4) node[dspnodeopen]{$\bm{Y}(\bm{x},s)$}; 
	
	\node at (10,4.5) {
		\parbox{3cm}{
			\centering $\bar{\bm{Y}}(s)$
	}};

	\node[draw, dspsquare, minimum width=1.2cm, minimum height=1cm](bk)at (8, 1) {
		\parbox{1.2cm}{
			\centering $-\Bs \Ks$
	}};
	\draw[dspline,double] (11,4) to (11,1) to (bk);
	\draw[dspline,double] (bk) to (4,1) to (4,3.4);
	
	\draw[dspline,double] (4,3.4) to (3.7,3.85);
	\node[draw,dspnodeopen](a) at (4,3.4){};
	\end{tikzpicture}
%	\vspace*{-2ex}
	\caption{State-space description of the 2D diffusion process described by \eqref{eq:p7}, \eqref{eq:p8} in the frequency domain with a generalized boundary term according to \eqref{eq:6} (switch open) and the desired boundary conditions \eqref{eq:p9} (switch closed) according to \eqref{eq:27}, \eqref{eq:30}.}
	\label{fig:3}
%	\vspace*{-3ex}
\end{figure}
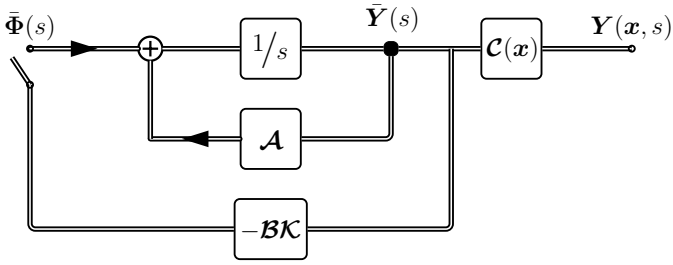

With the transformed boundary term in \eqref{eq:27} a modified state-space description (switch closed in Fig.~\ref{fig:3}) with a modified state matrix $\As_\mathrm{c}$ can be derived. Inserting \eqref{eq:27} into state equation \eqref{eq:15} leads to 
%\vspace*{-0.5ex}
\begin{align}
&s\bar{\bm{Y}}(s) = \As_\mathrm{c} \bar{\bm{Y}}(s) + \bar{\bm{y}}_\mathrm{i},  &\As_\mathrm{c} = \As - \Bs \Ks.
\label{eq:30}
\end{align} 
The modification of the state matrix $\As$ shifts the eigenvalues such that the realistic boundary conditions \eqref{eq:p9} are fulfilled. 
%These concepts are strongly related to the closed loop concept from control theory where a boundary control is designed to shift the system poles to a desired position \cite{Deutscher:IJC:2009,Willems:ICS:2007:4384643}. In this approach the poles are shifted by a boundary circuit, to fulfill a desired boundary behavior.

%\subsubsection{Time Domain Description}
Applying an inverse Laplace transform to state equation \eqref{eq:30} and output equation \eqref{eq:20} leads to a representation in terms of a matrix exponential
%\vspace*{-1.5ex} 
\begin{align}
&\bar{\bm{y}}(t) = e^{\As_\mathrm{c}t} \bar{\bm{y}}_\mathrm{i}, &\bm{y}(\bm{x},t) = \Cs(\bm{x}) \bar{\bm{y}}(t). 
\label{eq:31} 
\end{align}

%\vspace{-2.2ex}
\subsection{Simulation Algorithm}
\label{subsec:synth}

To obtain a simulation algorithm for the desired concentrations and fluxes in \eqref{eq:3}, the modified state-space description has to be transformed into the discrete-time domain. 
%\vspace*{-0.9ex}
\subsubsection{Impulse Invariant Transformation}
Sampling both equations in \eqref{eq:31} with the sampling interval $T$ as $t = kT$ and applying an impulse invariant transform (IIT) results in a synthesis algorithm in the discrete-time domain for the vector of variables 
\begin{align}
&\bar{\bm{y}}[k] = e^{\As_\mathrm{c}T}\bar{\bm{y}}[k-1] + \bar{\bm{y}}_\mathrm{i}\delta[k], &\bm{y}[\bm{x},k] = \Cs(\bm{x}) \bar{\bm{y}}[k].
\label{eq:32}
\end{align}
%\vspace*{-0.9ex}
\subsubsection{Transformation of Variables}
To obtain a simulation algorithm for the particle concentration $p_{\mathrm{r},z}(r,\varphi,z,t)$ in the receiver in Fig.~\ref{fig:1}, three steps are applied. First, the output equation of the state-space model is restricted to get the auxiliary concentration 
%\vspace{-1ex}
\begin{align}
q_{xy}[\bm{x},k] = \bm{c}_q^\tr(\bm{x})\,\bar{\bm{y}}[k],
\label{eq:33}
\end{align}
where $\bm{c}_q^\tr$ is the first row of transformation matrix $\Cs$ in \eqref{eq:21}. Second, 
the transformation of variables from Section~\ref{subsec:tov} turns the auxiliary concentration into
%\vspace{-1.8ex}
\begin{align}
&p_{xy}[\bm{x},k] = \exp\left(-\frac{u}{2}(y-y_0)-\frac{u^2}{4}kT\right) \cdot q_{xy}[\bm{x}, k]. 
\label{eq:34}
\end{align}
Third, the concentration in the cylindrical slice has to be recombined with the discrete-time equivalent of the concentration in a line segment \eqref{eq:p1a} to obtain a solution for the complete cylinder
%\vspace{-1ex}
\begin{align}
p_{\mathrm{r},z}[\bm{x},k] = p_{xy}[\bm{x},k] \cdot p_{d,z}[\bm{x},k],
\label{eq:35}
\end{align}
where the concentration $p_{xy}[\bm{x},k]$ can be calculated at any point of the system geometry by evaluation of  \eqref{eq:32}, \eqref{eq:33}, \eqref{eq:34}.

The individual steps, which are performed in the implementation and the underlying equations, are shown in Table~\ref{tab:1}. 
Steps 1-3 can be performed before the discrete-time simulation. 
In step 1 the sums in \eqref{eq:9} are truncated to $n = -N, \dots, N$ and $\mu = 0, \dots, M-1$. This leads to the total number of terms $Q = (2N + 1)\cdot M$. Step 3 constructs a state-space model of the diffusive channel with the generalized boundary term $\bar{\bm{\Phi}}$. In step 4 the desired boundary behavior is chosen - here \eqref{eq:p9} - and the feedback matrix $\Bs\Ks$ is constructed in step 5. With that matrix, the modified state matrix $\As_\mathrm{c}$ is computed in step 6. In step 7, the auxiliary concentration $q_{xy}$ is computed with \eqref{eq:32}, \eqref{eq:33} with a \texttt{for}-loop over time. Finally, the auxiliary concentration is transformed into the receiver concentration $p_{\mathrm{r},z}$ in steps 8,9.  

The algorithm assumes fixed parameters ($u$, $v_z$, etc.), but variable parameters can be easily incorporated and not all steps have to be repeated. Changing e.g. the velocity $u$, the algorithm stays the same for steps 1-3, a change of the horizontal flow $v_z$ affects only step 9. Also, to incorporate time-varying parameters, e.g. for a time-varying velocity $u$, steps 5,6 have to be moved inside the \texttt{for}-loop in step 7. This can affect the runtime of the algorithm.  
%\vspace*{-2.5ex}
\begin{table}[t]
	\caption{Individual steps of the proposed algorithm.}
	\label{tab:1}
%	\vspace*{-2.9ex}
	\small
	%	\resizebox{\columnwidth}{!}{%
	%	\small
	\renewcommand{\arraystretch}{1.1}
	\begin{tabular}{|p{5.9cm}|p{2cm}|}
		%	\hline
		%	Step & Equations\\
		%	\hline
		\hline
		Algorithm step & Equations \\
		\hline \hline
		1. Choose index range $N,M$ &  \\
		\hline
		2. Compute $M$ zeros of $J_n'(k_{\mu,n}a) = 0$ & \\
		\hline 
		3. Compute state and output matrices $\As, \Cs$ & \eqref{eq:12}, \eqref{eq:14}, \eqref{eq:19}, \eqref{eq:21}\\
		\hline \hline
		4. Choose desired boundary behavior & \eqref{eq:p9}\\
		\hline
		5. Compute feedback matrix $\Bs\Ks$ & \eqref{eq:27}, \eqref{eq:28}, \eqref{eq:29}\\
		\hline
		6. Compute modified state matrix $\As_\mathrm{c}$ & \eqref{eq:30}\\
		\hline \hline
		7. \texttt{for k = 1:simulationDuration} \phantom{aaaannaa} compute auxiliary concentration $q_{xy}$ & \eqref{eq:32}, \eqref{eq:33} \\ 
		\hline \hline
		8. Transformation of variables $q_{xy} \to p_{xy}$& \eqref{eq:34} \\
		\hline
		9. Compute receiver concentration $p_{xy} \to p_{\mathrm{r},z}$& \eqref{eq:35}\\
		\hline
	\end{tabular}
	%}
\end{table}
\renewcommand{\arraystretch}{1}

%%%% Experimental Verification %%%% 
%\vspace*{-4ex}
\section{Verification and Analysis}
\label{sec:experiments}

In this section, simulation results for the semi-analytical model described in Section~\ref{sec:model} are presented. The model outcome is compared to particle-based simulations \cite{Farsad_comprehensive_2016}. Furthermore, the proposed semi-analytical model is analyzed in view of runtime and accuracy. Also, the benefits of a semi-analytical model over purely numerical models are discussed.
%\vspace*{-1.8ex}
\subsection{Methods and Parameters}
\label{subsec:simbasis}
%\subsubsection{Geometrical and Physical Parameters}
%\label{subsubsec:param}
The radius of the cylinder is $a' = 1\cdot 10^{-4}\si{m}$. The diffusion is characterized by diffusion coefficient $D' = 1\cdot 10^{-10}\si{m^2/s}$ and horizontal uniform flow with $v'_z = 0.5\cdot 10^{-3}\si{m/s}$ (see \cite{Wicke_Molecular_2017}).
All parameters are normalized according to Section~\ref{subsec:norm} with reference length $\lambda = 1\cdot10^{-4}\si{m}$ and time $\tau = 1\cdot10^{2}\si{s}$. The injection of particles into the system is performed at $r_0 = 0.5\cdot a$, $\varphi_0 = -\pi$, $z = 0$. 

The simulations use a Matlab-Implementation of the semi-ana\-ly\-tical model with a total number of terms $Q=2870$ ($N=20$, $M = 70$). The presented results were produced by the algorithm in Table~\ref{tab:1} with a normalized sampling time $T = 1\cdot 10^{-4}$.

For validation, we employ a particle-based simulation of the advection-diffusion transport where the positions of $\Ntx$ particles are updated and tracked in discrete normalized time steps of length $\Delta t=0.5\cdot 10^{-3}$ following diffusion and drift, see e.g. \cite[Eq.~(1)]{Farsad_comprehensive_2016}.
Within each time step, if a particle crosses the channel boundary, it is reflected back into the channel.
For measuring the concentration within a line receiver, we consider a normalized cuboid of size $0.05\times 0.05\times 10$ approximating the line receiver in Fig.~\ref{fig:1}.
An estimate of the accumulated concentration is obtained by counting the number of observed particles within the cuboid and dividing by the surface area $(0.05)^2$ and the number of released particles $\Ntx$.
For $\Ntx=10^5$, results are averaged $100$ times.

The parameters, which are varied in the simulations, are velocity $u$ and the receiver position with a normalized length of $d = 10$. A total of four receiver positions are used with $\bm{x}_{1} = [r = 0.9a, \varphi = \frac{-\pi}{2}]$, $\bm{x}_{2} = [r = 0.9a, \varphi = \frac{-3\pi}{4}]$, each placed at $z_{1,2} = 75, 100$.
%\vspace*{-1.8ex}
\subsection{Particle-Based Verification}
\label{subsec:cs}

\begin{figure}
	\vspace{0.5ex}
    \centering
    \hspace*{-1.5ex}\includegraphics{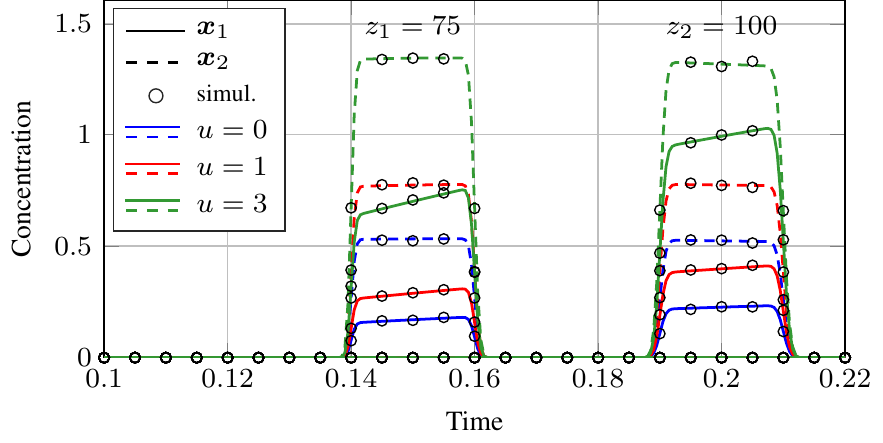}
%    \vspace*{-1.5ex}
    \caption{Received concentration $p_{\mathrm{r},z}(\bm{x},t)$ with \eqref{eq:35} (solid and dashed lines) and particle-based simulation (circle markers) for $u = 0,1,3$ (blue, red, green lines) at four receiver positions (see Section \ref{subsec:simbasis}).
    }
    \label{fig:6}
%    \vspace*{-4ex}
\end{figure}

In Fig.~\ref{fig:6}, the average concentration along the receiver (see Sec.~\ref{subsec:simbasis}) placed near the boundary of the cylinder is presented. The average concentration following an instantaneous point release of magnetic nano-particles represents the impulse response in a molecular communication system. For such a scenario, the increased signal strength induced by the drift velocity $u$ caused by a magnetic force is evaluated and can be directly related to the reliability of transmission \cite{Wicke_Molecular_2017}. The figure shows the result of the semi-analytical model according to \eqref{eq:35} (solid and dashed lines) and the outcome of the particle-based simulation (circle markers) for different drift velocities $u$. The semi-analytical model and the particle-based simulations are in excellent agreement.
The normalized width $t_s$ of the received signals in Fig.~\ref{fig:6} can be well approximated by $t_s \approx \nicefrac{d}{v_z} = 0.02$ which accounts for the length $d=10$ of the line receiver.
Hence, the diffusion along the $z$-axis does not significantly contribute to the width of the received signal.
In contrast, the height of the received signal critically depends on $u$, $z$, and the receiver position and is governed by the combination of cross-sectional diffusion and the vertical drift $u$.
%\vspace*{-1.8ex}
\subsection{Algorithm Analysis}
\label{subsec:ana}
\begin{figure}[t]
	\vspace{0.5ex}
	\begin{minipage}{0.45\linewidth}
		\hspace*{-1.2ex}
		\includegraphics{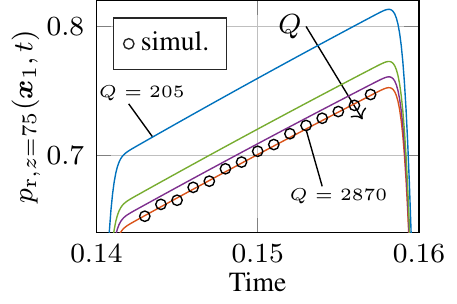}
	\end{minipage}\hfill
	\begin{minipage}{0.45\linewidth}
		\hspace*{-3ex}
		\includegraphics{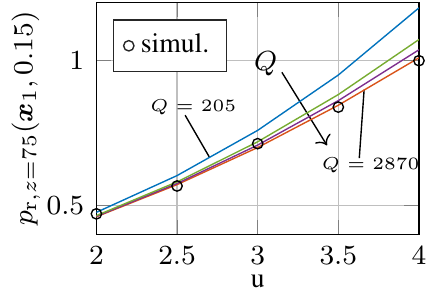}
	\end{minipage}
%	\vspace*{-1ex}
	\caption{Receiver concentration $p_{\mathrm{r},z=75}(\bm{x}_1,t)$ over time (left) and for increasing $u$ at $t = 0.15$ (right) with the proposed model for decreasing number of terms $Q=2870, 820, 410, 205$.}
	\label{fig:4}
	\vspace*{-2ex}
\end{figure}

%In this section the proposed algorithm is analyzed with respect to runtime, accuracy and complexity. Especially the trade-off between runtime and accuracy is discussed. 
%
Fig.~\ref{fig:4} shows two plots of the concentration $p_{\mathrm{r},z = 75}(\bm{x}_1,t)$ in the receiver simulated with the proposed model \eqref{eq:35} at the bottom of the cylinder. 
The left plot follows the blue line in Fig.~\ref{fig:6} and shows the particle concentration at $z = 75$ with a drift velocity of $u=3$. 
For the numerical analysis of the proposed model, the range of $n$ is fixed to $N = 20$ and the range of $\mu$ is varied as $M = 70, 20, 10, 5$. The circle markers are the result of the particle-based simulation. 
%
%The simulations are done for a decreasing number of terms $Q = N\cdot M$ (compare \eqref{eq:9}). The order range $n$ of the Bessel functions is fixed to $N = 41$ and the range of $\mu$ is varied $M = 70, 20, 10, 5$. The empty dots are the result of the particle based simulation as ground truth. 
%
The right plot in Fig.~\ref{fig:4} shows the receiver concentration $p_{\mathrm{r},z = 75}$ at time $t = 0.15$ as a function of $u$ for decreasing number of terms. 
As already shown in Section~\ref{subsec:cs} for $N=20, M =70$ ($Q = 2870$), the particle-based simulation and the results of the proposed semi-analytical are in excellent agreement. 

The runtime $t_Q$ of the algorithm is $t_{2870} = 125\,\si{s}$ to simulate a normalized duration of  $t_\mathrm{dur} = 0.5$ starting at $t = 0$.
For $Q = 820$ the semi-analytical model still agrees very well with the results of the particle-based model. The runtime decreases to $t_{820} = 4.5\,\si{s}$. 
A further reduction of the number of terms leads to visible deviations, but the runtime drops to $t_{410} = 0.9\,\si{s}$, $t_{210} = 0.5\,\si{s}$. 
For comparison, the runtime of the particle-based simulation is approx. $t_\mathrm{p} \approx 30\,\si{min}$ to simulate a normalized duration $t_\mathrm{dur} = 0.5$. 
The results in Fig.~\ref{fig:4} reveal a trade-off between runtime and accuracy. This shows the flexibility of the proposed model: 
If the exact concentrations are needed, the number of terms, $Q$, can be chosen large at the cost of runtime. On the other hand, to study the rough behavior of a channel for different parameter sets, the number of terms can be reduced to run many simulations in a short time.
The plot on the right hand side of Fig.~\ref{fig:4} shows that the accuracy also depends on the velocity $u$ and, for a given $Q$, decreases for increasing $u$. 
%For an analysis of the complexity of the algorithm, the most interesting step is the iteration over time in step 7 (table~\ref{tab:1}), which depends on the programming language. Using a representation of \eqref{eq:33} in terms of sums \eqref{eq:9} leads to a nesting of three \texttt{for}-loops, which leads to a complexity of $\mathcal{O}(n^3)$. Using a vector-based implementation of \eqref{eq:33} reduces the number of \texttt{for}-loops and results in a complexity of $\mathcal{O}(n)$.

%\vspace*{-1.8ex}
\subsection{Benefits of the Semi-Analytical Model}

The proposed semi-analytical state-space model (see Section~\ref{sec:model}) offers many advantages over purely numerical models. 

In particular, the semi-analytical description of the diffusion process allows to compute the channel impulse response in an analytical manner.
Hence, solutions in the form of \eqref{eq:32} can be related to parameters relevant for communication system design (e.g. inter-symbol interference) more directly than solutions for purely numerical models. 
Additionally, the complexity of the simulation algorithm is flexible (varying $M,N$), 
and can be adjusted to 
%sdepending on 
the simulation purpose (see Section~\ref{subsec:ana}). 
Knowing the eigenvalues $s_{\mu,n}$ of the diffusion process provides insight into the behavior, e.g. damping or asymptotic behavior. As the output variable $\bm{y}$ in \eqref{eq:32} also contains the fluxes $\bm{i}$, they can be obtained without any additional effort.

The formulation of the solution in terms of a state-space representation in Section \ref{subsubsec:ssd} with the general boundary term $\bar{\bm{\Phi}}(s)$ can be extended to different kinds of boundary conditions. In this paper, the techniques from Section~\ref{subsec:fcl} are used to realize the boundary conditions \eqref{eq:p9}. 
\cbstart
By the same technique also other types of boundary behavior can be realized, e.g. semi-permeable membranes or non-reflective cell walls \cite{lnt2017-47}. The incorporation of different boundary conditions can be seen as an extension of the general state-space description, so the computation of the general model (Steps 1-3 in Table \ref{tab:1}) remains the same. 
\cbend
This allows the simulation of complex diffusive systems/channels in a semi-analytical manner by separating the general model from the desired boundary behavior. 

Also, the interconnection of several semi-analytical models for larger biochemical structures is possible by the techniques shown in Section~\ref{subsec:fcl} \cite{Willems:ICS:2007:4384643}. E.g. several cylindrical models can be used to model a cascade of blood vessels, or several spherical models can realize complex interacting cell structures.

%%%% Analysis %%%% 
%\input{analysis}

%%%% Conclusion & Further Works %%%% 
%\vspace*{-1.5ex}
\section{Conclusion and Future Work}
\label{sec:conclusion}

In this paper, we presented a semi-analytical model for the diffusion of particles influenced by a vertical force and horizontal uniform drift in a cylindrical shape modeling e.g. the transport of information carrying magnetic nano-particles within a blood vessel under the influence of a magnetic field. 
The semi-analytical model is based on a modal expansion of spatial differential operators, and is useful e.g., for studying the signal strength of the channel impulse response.
%and a representation in terms of a state-space description. 
After the design of a general model for the diffusion process, the boundary behavior of the system is adjusted by a feedback loop. The validity of the model is verified by comparisons to particle-based simulation results. An analysis of the algorithm reveals several benefits (e.g. in runtime and flexibility) compared to purely numerical models.
%
%and the accuracy and runtime of the proposed algorithm is discussed.   

Future work might investigate extensions to non-uniform laminar flow, more complex receiver models, and different geometries. 
\cbstart
Based on the time-variant source signals in \cite{lnt2017-27}, the excitation of the model by realistic time- and space-variant source signals will be considered.
\cbend

%\footnotesize This work has been supported by the German Research Foundation (Deutsche Forschungsgemeinschaft, DFG),  grant number RA 801/6-1, SCHO 831/6-1.

%\footnotesize(This work has been supported by the German Research Foundation (Deutsche Forschungsgemeinschaft, DFG) under grant number RA 801/6-1)

\bibliographystyle{IEEEtran}
{\footnotesize
	\bibliography{./../Bib/nanoCom18}
}

\end{document}